\documentclass[10pt,conference]{IEEEtran}
\usepackage{maxim}
\usepackage{graphicx}
\usepackage{subfigure}

\def\cA{{\cal A}}

\def\cF{{\cal F}}

\def\cS{{\cal S}}

\def\cX{{\cal X}}

\def\hD{\widehat{D}}
\def\hcX{\widehat{\cal X}}
\def\hX{\widehat{X}}
\def\hx{\widehat{x}}

\def\wh#1{\widehat{#1}}
\def\td#1{\widetilde{#1}}

\begin{document}

\title{Joint Universal Lossy Coding and Identification\\
of I.I.D. Vector Sources}

\author{\authorblockN{Maxim Raginsky}
\authorblockA{Beckman Institute and the University of Illinois\\
405 N Mathews Ave, Urbana, IL 61801, USA \\
Email: maxim@uiuc.edu}}
\maketitle

\begin{abstract}
The problem of joint universal source coding and modeling, addressed by Rissanen in the context of lossless codes, is generalized to fixed-rate lossy coding of continuous-alphabet memoryless sources. We show that, for bounded distortion measures, any compactly parametrized family of i.i.d. real vector sources with absolutely continuous marginals (satisfying appropriate smoothness and Vapnik--Chervonenkis learnability conditions) admits a joint scheme for universal lossy block coding and parameter estimation, and give nonasymptotic estimates of convergence rates for distortion redundancies and variational distances between the active source and the estimated source. We also present explicit examples of parametric sources admitting such joint universal compression and modeling schemes.
\end{abstract}

\section{Introduction}
\label{sec:intro}

In universal data compression, a single code achieves asymptotically optimal performance on all sources within a given family. Intuition suggests that a good universal coder should acquire an accurate model of the source statistics from a sufficiently long data sequence and incorporate this knowledge in its operation. For lossless codes, this intuition has been made rigorous by Rissanen \cite{Ris84}. Under his scheme, the data are encoded in a {\em two-stage} set-up, in which the binary representation of each source block consists of two parts: (1) a suitably quantized maximum-likelihood estimate of the source parameters, and (2) lossless encoding of the data matched to the acquired model; the redundancy of the resulting code converges to zero as $\log n/n$, where $n$ is the block length.

In this paper, we extend Rissanen's idea to {\em lossy} block coding (vector quantization) of i.i.d. sources with values in $\R^d$ for some finite $d$. Specifically, let $\{X_i\}^\infty_{i=-\infty}$ be an i.i.d. source with the marginal distribution of $X_1$ belonging to some indexed class $\{P_\theta : \theta \in \Theta\}$ of absolutely continuous distributions on $\R^d$, where $\Theta$ is a bounded subset of $\R^k$ for some $k$. For bounded distortion measures, our main result, Theorem~\ref{thm:wu}, states that if the class $\{P_\theta\}$ satisfies certain smoothness and learnability conditions, then there exists a sequence of finite-memory lossy block codes that achieves asymptotically optimal compression of each source in the class and permits asymptotically exact identification of the active source with respect to the {\em variational distance}, defined as $d_V(P,Q) \deq \sup_B |P(B) - Q(B)|$, where the supremum is over all Borel subsets of $\R^d$. The overhead rate and the distortion redundancy of the scheme converge to zero as $O(\log n/n)$ and $O(\sqrt{\log n/n})$, respectively, where $n$ is the block length, while the active source can be identified up to a variational ball of radius $O(\sqrt{\log n/n})$ eventually almost surely. We also describe an extension of our scheme to unbounded distortion measures satisfying a certain moment condition, and present two examples of parametric families satisfying the regularity conditions of Theorem~\ref{thm:wu}.

While most existing schemes for universal lossy coding rely on {\em implicit} identification of the active source (e.g., through topological covering arguments \cite{NeuGraDav75}, Glivenko--Cantelli uniform laws of large numbers \cite{LinLugZeg94}, or nearest-neighbor code clustering \cite{ChoEffGra96}), our code builds an {\em explicit model} of the mechanism responsible for generating the data and then selects an appropriate code for the data on the basis of the model. This ability to simultaneously model and compress the data may prove useful in such applications as {\em media forensics} \cite{MouKoe05}, where the parameter $\theta$ could represent evidence of tampering, and the aim is to compress the data in such a way that the evidence can be later extracted with high fidelity from the compressed version. Another key feature of our approach is the use of Vapnik--Chervonenkis theory \cite{DevLug01} in order to connect universal encodability of a class of sources to the combinatorial ``richness" of a certain collection of decision regions associated with the sources. In a way, Vapnik--Chervonenkis estimates can be thought of as an (imperfect) analogue of the combinatorial method of types for finite alphabets \cite{CsiKor81}.

\section{Preliminaries}
\label{sec:prelims}

Let $\{X_i\}^\infty_{i=-\infty}$ be an i.i.d. source with alphabet $\cX$, such that the marginal distribution of $X_1$ comes from an indexed class $\{P_\theta : \theta \in \Theta\}$. For any $t \in \Z$ and any $m,n \ge 0$, let $X^n_m(t)$ denote the segment $(X_{tn-m+1},X_{tn-m+2},\cdots,X_{tn})$ of $\{X_i\}$, with $X^n_0(t)$ understood to denote an empty string for all $n,t$. We shall abbreviate $X^n_n(t)$ to $X^n(t)$.

Consider coding $\{X_i\}$ into the reproduction process $\{\hX_i\}$ with alphabet $\hcX$ by means of a stationary lossy code with block length $n$ and memory length $m$ [an $(n,m)$-block code, for brevity]. Such a code consists of an encoder $\map{f}{\cX^n \times \cX^m}{\cS}$ and a decoder $\map{\phi}{\cS}{\hcX^n}$, where $\cS$ is a collection of fixed-length binary strings: $\hX^n(t) = \phi(f(X^n(t),X^n_m(t-1))), \forall t \in \Z$. That is, the encoding is done in blocks of length $n$, but the encoder is also allowed to observe a fixed finite amount of past data. Abusing notation, we shall denote by $C^{n,m}$ both the composition $\phi \circ f$ and the encoder-decoder pair $(f,\phi)$; when $m=0$, we shall use a more compact notation $C^n$. The number $R(C^{n,m}) \deq n^{-1}\log |\cS|$ is the rate of $C^{n,m}$ in bits per letter. The distortion between the source $n$-block $X^n = (X_1,\cdots,X_n)$ and its reproduction $\hX^n = (\hX_1,\cdots,\hX_n)$ is given by $\rho(X^n,\hX^n) = \sum^n_{i=1}\rho(X_i,\hX_i)$, where $\map{\rho}{\cX \times \hcX}{\R^+}$ is a single-letter distortion measure.

Suppose $X_1 \sim P_\theta$ for some $\theta \in \Theta$. Since the source is i.i.d., and the code $C^{n,m}$ does not vary with time, the process $\{(X_i,\hX_i)\}$ is $n$-stationary, and the average distortion of $C^{n,m}$ is $D_\theta(C^{n,m}) = n^{-1}\E_\theta\big[\rho(X^n(1),\hX^n(1))\big]$, where $\hX^n(1) \equiv \phi(f(X^n(1),X^n_m(0)))$. The optimal performance achievable on $P_\theta$ by any finite-memory $n$-block code at rate $R$ is given by the $n$th-order operational distortion-rate function (DRF)
$$
\hD^{n,*}_\theta(R) \deq \inf_{m \ge 0} \,\, \inf_{C^{n,m}: R(C^{n,m}) \le R} \,\, D_\theta(C^{n,m}),
$$
where the asterisk denotes the fact that we allow any finite memory length. For zero-memory $n$-block codes the corresponding DRF is
$$
\hD^n_\theta(R) \deq \Inf_{C^n: R(C^n) \le R} D_\theta(C^n).
$$
Clearly, $\hD^{n,*}_\theta(R) \le \hD^n_\theta(R)$. Conversely, we can use memoryless minimum-distortion encoders to convert any $(n,m)$-block code into a zero-memory $n$-block code without increasing either distortion or rate, so $\hD^n_\theta(R) = \hD^{n,*}_\theta(R)$. Finally, the best performance achievable by any block code, with or without memory, is given by the operational DRF $\hD_\theta(R) \deq \Inf_{n \ge 1} \hD^n_\theta(R) = \Lim_{n \to \infty} \hD^n_\theta(R)$; since the source is i.i.d., $\hD_\theta(R)$ is equal to the Shannon DRF $D_\theta(R)$ by the source coding theorem and its converse.

We are interested in sequences of codes that asymptotically achieve optimal performance across the entire class $\{P_\theta \}$. Let $\{C^{n,m}\}^\infty_{n=1}$ be a sequence of $(n,m)$-block codes, where the memory length $m$ may depend on $n$, such that $R(C^{n,m}) \to R$. Then $\{C^{n,m}\}$ is {\em weakly minimax universal} for $\{P_\theta\}$ if the {\em distortion redundancy} $\delta_\theta(C^{n,m}) \deq D_\theta(C^{n,m}) - D_\theta(R)$ converges to zero as $n \to \infty$ for every $\theta \in \Theta$.\footnote{See \cite{NeuGraDav75} for other notions of universality for source codes.} We shall follow Chou {\em et al.} \cite{ChoEffGra96} and split $\delta_\theta(C^{n,m})$ into two terms:
$$
\delta_\theta(C^{n,m}) = \big(D_\theta(C^{n,m}) - \hD^n_\theta(R)\big) + \big(\hD^n_\theta(R) - D_\theta(R)\big).
$$
The first term, which we shall call the $n$th-order redundancy and denote by $\delta^n_\theta(C^{n,m})$, is the excess distortion of $C^{n,m}$ relative to the best $n$-block code for $P_\theta$, while the second term gives the extent to which the best $n$-block code falls short of the Shannon optimum. Note that $\delta_\theta(C^{n,m}) \to 0$ if and only if $\delta^n_\theta(C^{n,m}) \to 0$, since $\hD^n_\theta(R) \to D_\theta(R)$ by the source coding theorem.

\section{Informal description of the system}
\label{sec:system}

As stated in the Introduction, we are after a sequence of lossy block codes that would not only be universally optimal for a given class $\{P_\theta\}$ of i.i.d. sources with values in $\R^d$, but would also permit asymptotically reliable identification of the source parameter $\theta \in \Theta$. We formally state and prove our result in Section~\ref{sec:mainthm}; here we outline the main idea behind it.

Fix the block length $n$, and denote by $X^n$ the current $n$-block $X^n(t)$ and by $Z^n$ the preceding $n$-block $X^n(t-1)$. Let us assume that we can find for each $\theta \in \Theta$ an $n$-block code $C^n_\theta$ at the desired rate $R$ which achieves the $n$th-order DRF for $P_\theta$: $D_\theta(C^n_\theta) = \hD^n_\theta(R)$. The basic idea is to construct an $(n,n)$-block code $C^{n,n}$ that first estimates the parameter $\theta$ of the active source from $Z^n$ and then codes $X^n$ with the code $C^n_{\wh{\theta}(Z^n)}$, where $\wh{\theta}(Z^n)$ is the estimate of $\theta$, suitably quantized.

Suppose the encoder can use $Z^n$ to identify the active source up to a variational ball of radius $O(\sqrt{\log n/n})$. Next, suppose that the parameters of the estimated source (assumed to belong to a bounded subset of $\R^k$ for some $k$) are quantized to $O(\log n)$ bits in such a way that the variational distance between any two sources whose parameters lie in the same quantizer cell is $O(\sqrt{1/n})$. If $\wh{\theta} = \wh{\theta}(Z^n)$ is the quantized parameter estimate, then the variational distance between $P_{\wh{\theta}}$ and the ``true" source $P_\theta$ is $O(\sqrt{\log n/n})$, which for bounded distortion functions implies an $O(\sqrt{\log n/n})$ upper bound on the distortion redundancy $\delta^n_\theta(C^n_{\wh{\theta}})$.

More formally, let $\map{\td{f}}{\cX^n}{\td{\cS}}$ denote the map that sends each $Z^n$ to the binary representation of $\wh{\theta}(Z^n)$, where $\td{\cS}$ is a collection of fixed-length binary strings with $\log |\td{\cS}| = O(\log n)$, and let $\map{\td{\psi}}{\td{\cS}}{\Theta}$ be the parameter decoder that maps each $\td{s} \in \td{\cS}$ to its reproduction: $\wh{\theta}(Z^n) = \td{\psi}(\td{f}(Z^n))$. Thus, to each $\td{s} \in \td{\cS}$ there corresponds an $n$-block code $C^n_{\td{\psi}(\td{s})}$, which we denote more compactly by $C^n_{\td{s}} = (f_{\td{s}},\phi_{\td{s}})$. Our $(n,n)$-block code $C^{n,n}$ thus has the encoder $f(X^n,Z^n) \deq \td{f}(Z^n)f_{\td{f}(Z^n)}(X^n)$ and the corresponding decoder $\phi(\td{f}(Z^n)f_{\td{f}(Z^n)}(X^n)) \deq \phi_{\td{f}(Z^n)}(f_{\td{f}(Z^n)}(X^n))$. That is, the binary string emitted by the encoder consists of two parts: (a) the {\em header} containing a binary description of the chosen code [equivalently, of the estimated source $P_{\wh{\theta}(Z^n)}$], and (b) the {\em body} containing the binary description of the data $X^n$ using the chosen code at rate $R$. The combined rate of $C^{n,n}$ is $R + n^{-1} \log |\td{\cS}| = R + O(\log n/n)$ bits per letter, while the expected distortion with respect to $P_\theta$ is
\begin{eqnarray}
D_\theta(C^{n,m}) &=& \frac{1}{n}\E_\theta\left[\rho(X^n,\phi(f(X^n,Z^n)))\right] \nonumber \\
&=& \frac{1}{n} \E_\theta\left\{ \E_\theta\big[ \rho(X^n,C^n_{\td{f}(Z^n)}(X^n))\big|Z^n\big]\right\} \nonumber \\
&=& \E_\theta\big[D_\theta(C^n_{\td{f}(Z^n)})\big].
\label{eq:2stdist_mem}
\end{eqnarray}
This scheme is universal because the map $\td{f}$ and the subcodes $C^n_{\td{s}}$ are chosen so that $\E_\theta\big[D_\theta(C^n_{\td{f}(Z^n)})\big] - \wh{D}^n_\theta(R) = O(\sqrt{\log n/n})$ for each $\theta \in \Theta$. Note that decoder can not only decode the data in a near-optimal fashion, but also  identify the active source up to a variational ball of radius $O(\sqrt{\log n/n})$.
  
\begin{figure*}
\centerline{
\subfigure[Encoder]{\includegraphics[width=2.75in]{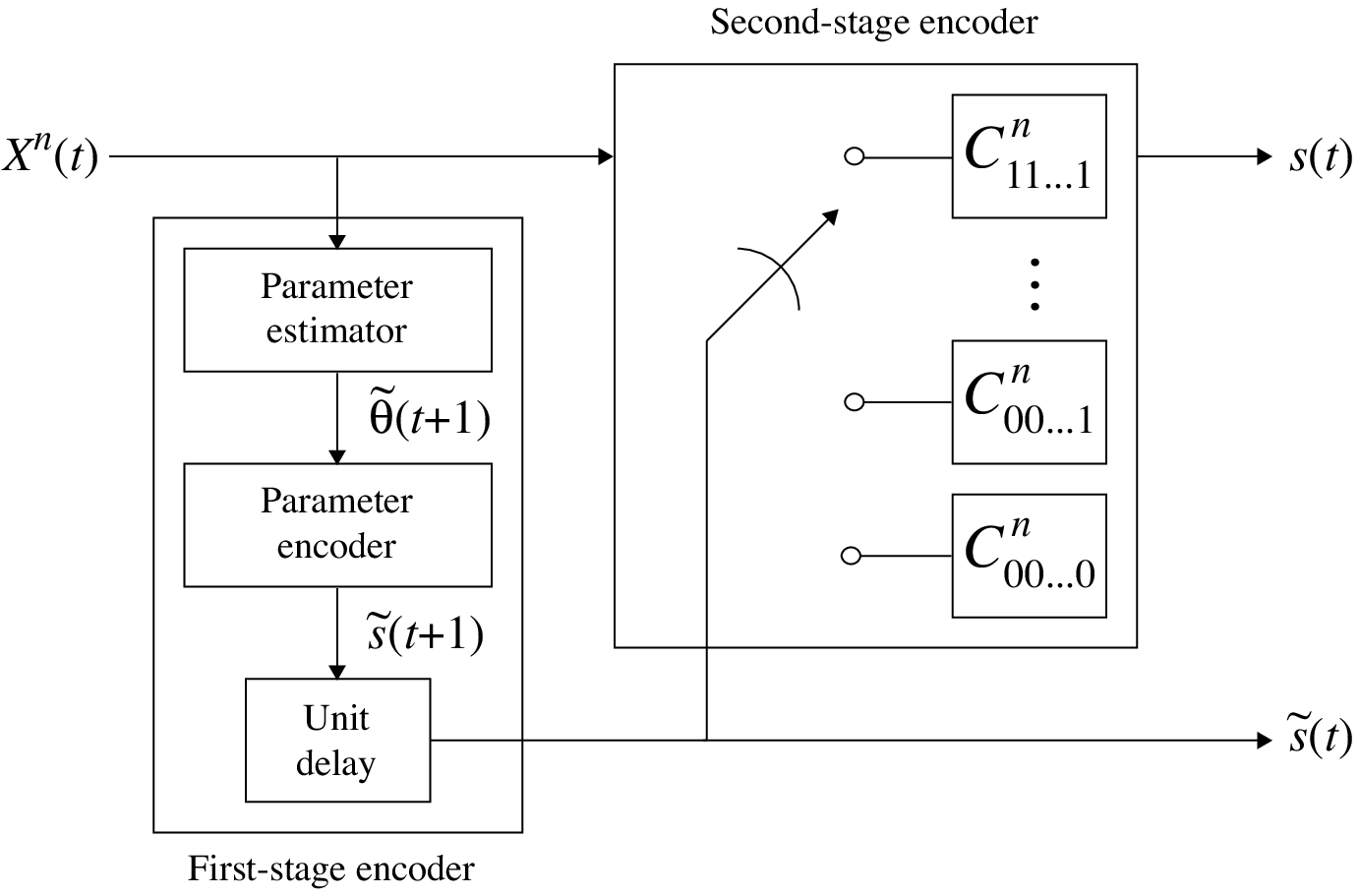}\label{fig:encoder}
}
\hfil
\subfigure[Decoder]{\includegraphics[width=2.75in]{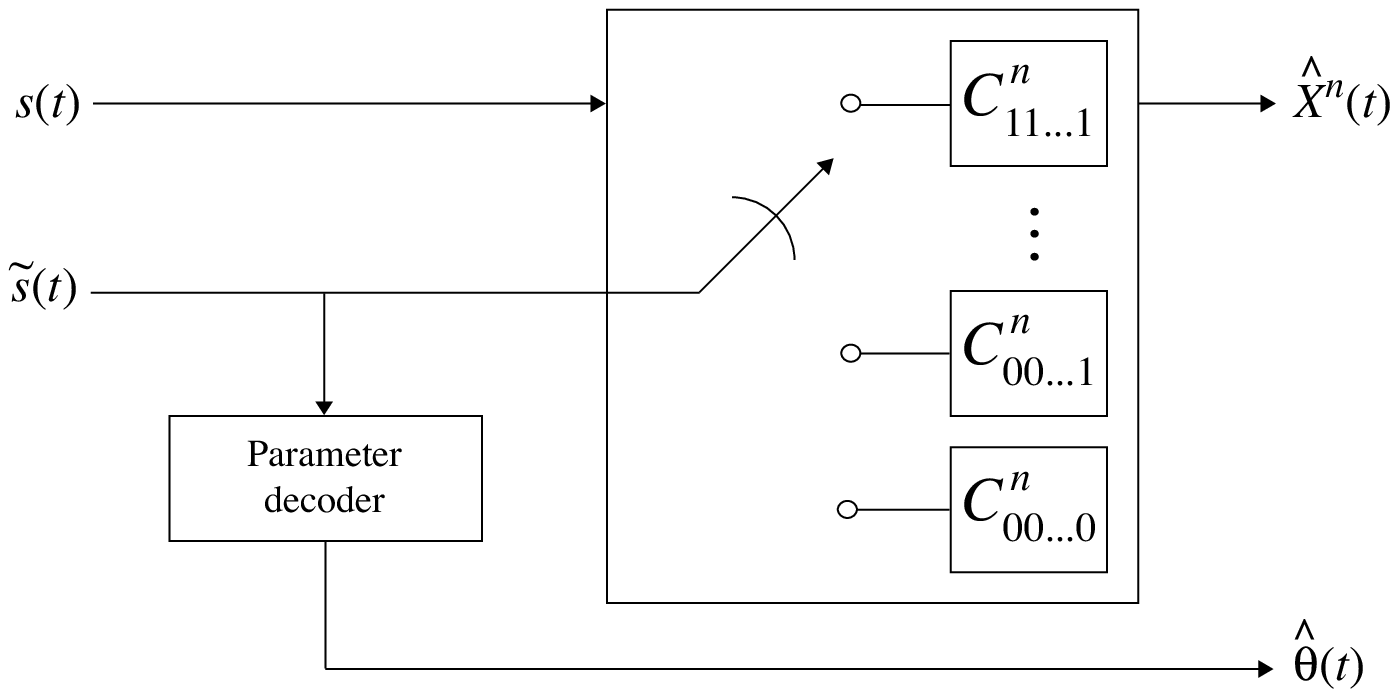}\label{fig:decoder}
}}
\caption{The two-stage scheme for joint universal lossy source coding and identification.}
\label{fig:twostage}
\end{figure*}

We remark that our scheme is a modification of the two-stage code of Chou {\em et al.} \cite{ChoEffGra96}, the difference being that here the subcode $C^n_{\td{s}}$, used to encode the current $n$-block $X^n$, is selected on the basis of the preceding $n$-block $Z^n$. Nonetheless, we shall adopt the terminology of \cite{ChoEffGra96} and refer to $\td{f}$ as the {\em first-stage encoder}. The structure of the encoder and the decoder in our scheme is displayed in Fig.~\ref{fig:twostage}.

\section{The main theorem}
\label{sec:mainthm}

Before stating and proving our result, let us list our assumptions, as well as fix some auxiliary results and notation.

\noindent{\bf The source models.} Let $\{X_i\}^\infty_{i=-\infty}$ be an i.i.d. source with alphabet $\cX$ and with the marginal distribution of $X_1$ belonging to an indexed class $\{P_\theta : \theta \in \Theta\}$, such that the following conditions are satisfied:
\newcounter{Lcount}
\begin{list}{(S.\arabic{Lcount})}{\usecounter{Lcount}}
\item $\cX$ is a measurable subset of $\R^d$.\label{src:alphabet}
\item Each $P_\theta$ is absolutely continuous with pdf $p_\theta$.
\end{list}
\setcounter{Lcount}{0}

\noindent{\bf Distortion function.} The distortion function $\rho$ is assumed to satisfy the following requirements:
\begin{list}{(D.\arabic{Lcount})}{\usecounter{Lcount}}
\item $\Inf_{\hx \in \hcX} \rho(x,\hx) = 0$ for all $x \in \cX$.
\item $\Sup_{x \in \cX, \hx \in \hcX} \rho(x,\hx) = K < \infty$.
\end{list}
Under these conditions, it can be proved (in a manner similar to the proof of Thm.~2 of Linder {\em et al.} \cite{LinLugZeg94}) that
\begin{equation}
\hD^n_\theta(R) = D_\theta(R) + O(\sqrt{\log n/n})
\label{eq:drf_conv}
\end{equation}
for each $\theta \in \Theta$ and for all rates $R$ such that $D_\theta(R) > 0$ (this condition is automatically satisfied in our case since all the $P_\theta$'s are absolutely continuous). The constant implicit in $O(\cdot)$ depends on both $\theta$ and $R$.

\noindent{\bf Vapnik--Chervonenkis theory.} Given a collection $\cA$ of measurable subsets of $\R^d$, its {\em Vapnik-Chervonenkis (VC) dimension} $V(\cA)$ is defined as the largest integer $n$ for which
\begin{equation}
\max_{x^n \in (\R^d)^n} |\{(1_{\{x_1 \in A\}},\cdots,1_{\{x_n \in A\}}) : A \in \cA \} | = 2^n;
\label{eq:vc}
\end{equation}
if (\ref{eq:vc}) holds for all $n$, then $V(\cA) = \infty$. If $V(\cA) < \infty$, we say that $\cA$ is a VC class. For any such class, one can give finite-sample bounds on uniform deviations of probabilities of events in that class from their relative frequencies. That is, if $X^n = (X_1,\cdots,X_n)$ is an i.i.d. sample from a distribution $P$, and if $\cA$ is a VC class with $V(\cA) \ge 2$, then
$$
\Pr \left\{\sup_{A \in \cA} |P_{X^n}(A) - P(A)| > \epsilon\right\} \le 8n^{V(\cA)}e^{-n\epsilon^2/32}, \forall \epsilon > 0
$$
and
$$
\E\left\{\sup_{A \in \cA} |P_{X^n}(A) - P(A)| \right\} \le c\sqrt{\log n/n},
$$
where $P_{X^n}$ is the empirical distribution of $X^n$ and the constant $c$ depends on $V(\cA)$ but not on $P$.\footnote{Using more refined techniques, the $c\sqrt{\log n/n}$ bound can be improved to $c'/\sqrt{n}$, where $c'$ is another constant, but $c'$ is much larger than $c$, so any benefit of the new bound shows only for ``impractically" large values of $n$.} (See, e.g., \cite{DevLug01} for details.)

\begin{theorem}\label{thm:wu} Let $\{X_i\}^\infty_{i=-\infty}$ be an i.i.d. source satisfying Conditions (S.1) and (S.2), and let $\rho$ be a distortion function satisfying Conditions (D.1) and (D.2). Assume the following:
\begin{enumerate}
\item $\Theta$ is a bounded subset of $\R^k$ for some $k$.
\item The map $\theta \mapsto P_\theta$ is {\em uniformly locally Lipschitz}: there exist constants $r,\beta > 0$ such that, for each $\theta \in \Theta$, $d_V(P_\theta,P_\eta) \le \beta \| \theta - \eta \|$ for all $\eta \in B_r(\theta)$, where $\| \cdot \|$ is the Euclidean norm on $\R^k$ and $B_r(\theta)$ is an open ball of radius $r$ centered at $\theta$.
\item The collection $\cA_\Theta$ of all sets of the form $A_{\theta,\eta} = \{x \in \cX: p_\theta(x) > p_\eta(x) \}$ with $\theta \neq \eta$ (the so-called {\em Yatracos class} associated with $\{P_\theta\}$ \cite{Yat85,DevLug96,DevLug97}) is a VC class.
\end{enumerate}
Also, suppose that for each $n,\theta$ there exists an $n$-block code $C^n_\theta = (f_\theta,\phi_\theta)$ at rate of $R$ bits per letter achieving the $n$th-order operational DRF for $\theta$: $D_\theta(C^n_\theta) = \hD^n_\theta(R)$. Then there exists an $(n,n)$-block code $C^{n,n}$ with
\begin{equation}
R(C^{n,n}) = R + O(\log n/n),
\label{eq:rate}
\end{equation}
such that for every $\theta \in \Theta$
\begin{equation}
\delta_\theta(C^{n,n}) = O(\sqrt{\log n/n}).
\label{eq:distred}
\end{equation}
Therefore, the sequence of codes $\{C^{n,n}\}^\infty_{n=1}$ is weakly minimax universal for $\{P_\theta : \theta \in \Theta\}$ at rate $R$. Furthermore, for each $n$ the first-stage encoder $\td{f}$ and the corresponding parameter decoder $\td{\psi}$ are such that
\begin{equation}
d_V(P_\theta,P_{\td{\psi}(\td{f}(X^n))}) = O(\sqrt{\log n/n}) \qquad \mbox{$P_\theta$-a.s.}
\label{eq:srcident}
\end{equation}
The constants implicit in the $O(\cdot)$ notation in (\ref{eq:rate}) and (\ref{eq:srcident}) are independent of $\theta$.
\end{theorem}

\begin{proof} The proof is by construction of a two-stage $(n,n)$-block code as outlined in Sec.~\ref{sec:system}. As before, let $X^n$ denote the current $n$-block $X^n(t)$, and let $Z^n$ be the preceding $n$-block $X^n(t-1)$. We first define our first-stage encoder $\td{f}$, which we shall realize as the composition $\td{g} \circ \td{\theta}$ of a parameter estimator $\td{\theta}$ and a lossy parameter encoder $\td{g}$ (cf.~Fig.~\ref{fig:twostage}). For any $z^n \in \cX^n$ and for any $\theta \in \Theta$, let $\Delta_\theta(z^n) \deq \Sup_{A \in \cA_\Theta} |P_\theta(A) - P_{z^n}(A)|$, where $P_\theta(A) \equiv \int_A p_\theta(x)dx$. Define $\td{\theta}(z^n)$ as any $\theta^* \in \Theta$ such that $\Delta_{\theta^*}(z^n) < \Inf_{\theta \in \Theta}\Delta_\theta(z^n) + 1/n$, where the extra $1/n$ ensures that at least one such $\theta^*$ exists. The map $z^n \mapsto \td{\theta}(z^n)$ is the so-called {\em minimum-distance density estimator} of Devroye and Lugosi \cite{DevLug96, DevLug97}, which satisfies
\begin{equation}
d_V(P_\theta,P_{\td{\theta}(Z^n)}) \le 2\Delta_\theta(Z^n) + 3/2n.
\label{eq:mindist}
\end{equation}
Since $\cA_\Theta$ is a VC class,
\begin{equation}
\E_\theta[d_V(P_\theta,P_{\td{\theta}(Z^n)})] \le c\sqrt{\log n/n} + 3/2n,
\label{eq:avmindist}
\end{equation}
for each $\theta \in \Theta$, where $c > 0$ depends on $V(\cA_\Theta)$.

Next, we construct the lossy encoder $\td{g}$. Since $\Theta$ is bounded, it is contained in some cube $M$ of side $J \in \N$. Let $\{M^{(n)}_1,M^{(n)}_2,\cdots,M^{(n)}_K\}$ be a partitioning of $M$
into contiguous cubes of side $1/\lceil n^{1/2} \rceil$, so
that $K \le (Jn^{1/2})^k$. Represent each $M^{(n)}_j$ that intersects
$\Theta$ by a unique fixed-length binary string $\td{s}_j$, and let
$\td{\cS} = \{\td{s}_j\}$.  Then if a given $\theta \in \Theta$ is
contained in $M^{(n)}_j$, map it to $\td{s}_j$, $\td{g}(\theta) =
\td{s}_j$; this can be described by a string of no more than $k(\log n^{1/2} + \log J)$
bits. For each $M^{(n)}_j$ that intersects
$\Theta$, choose a reproduction $\wh{\theta}_j \in M^{(n)}_j \cap
\Theta$ and designate $C^n_{\wh{\theta}_j}$ as the corresponding $n$-block code $C^n_{\td{s}_j}$. The parameter decoder
$\map{\td{\psi}}{\td{\cS}}{\Theta}$ is then given by $\td{\psi}(\td{s}_j) = \wh{\theta}_j$.

The rate of the resulting $(n,n)$-block code $C^{n,n}$ does not exceed $R + n^{-1}k(\log n^{1/2} + \log J)$ bits per letter, which proves (\ref{eq:rate}). By (\ref{eq:2stdist_mem}), the average distortion of $C^{n,n}$ on the source
$P_\theta$ is given by $D_\theta(C^{n,n}) = \E_\theta \big[ D_\theta\big(C^n_{\wh{\theta}}
  \big) \big]$, where $\wh{\theta} = \wh{\theta}(Z^n) \equiv
\td{\psi}(\td{f}(Z^n))$. From standard quantizer mismatch arguments and the triangle inequality, 
$$
\delta^n_\theta(C^n_{\wh{\theta}}) \le 4K[
  d_V(P_\theta,P_{\td{\theta}}) +
  d_V(P_{\td{\theta}},P_{\wh{\theta}})].
$$
Taking
expectations, we get
\begin{equation} \delta^n_\theta(C^{n,n}) \le 4K
\big\{\E_\theta [d_V(P_\theta,P_{\td{\theta}})] + \E_\theta
  [d_V(P_{\td{\theta}},P_{\wh{\theta}})]\big\}.
\label{eq:distbound}
\end{equation}
We now estimate separately each term in the curly brackets in (\ref{eq:distbound}). Using (\ref{eq:avmindist}), we can bound the first term by
\begin{equation}
\E_\theta [d_V(P_\theta,P_{\td{\theta}})] \le c \sqrt{\log n/n}
+ 3/2n.
\label{eq:1st_term}
\end{equation}
The second term involves $\td{\theta} = \td{\theta}(Z^n)$ and its quantized version
$\wh{\theta}$, where $\| \td{\theta} - \wh{\theta} \| \le
\sqrt{k/n}$ by construction of $\td{g},\td{\psi}$. Using the assumption that the map $\theta \mapsto P_\theta$ is uniformly locally Lipschitz, as well as the fact that $d_V(P,Q) \le 1$ for any two distributions $P,Q$, it is not hard to show that there exists a constant $\beta'$ such that
\begin{equation}
d_V(P_{\td{\theta}},P_{\wh{\theta}})
\le \beta' \sqrt{k/n}
\label{eq:2nd_term_noexp}
\end{equation}
and consequently that
\begin{equation}
\E_\theta [d_V(P_{\td{\theta}},P_{\wh{\theta}})] \le \beta'\sqrt{k/n}.
\label{eq:2nd_term}
\end{equation}
Substituting the bounds (\ref{eq:1st_term}) and (\ref{eq:2nd_term})
into (\ref{eq:distbound}) yields
$$
\delta^n_\theta(C^{n,n}) \le K \left(
  4c\sqrt{\log n/n} + 6/n + 4\beta' \sqrt{k/n}
  \right),
$$
whence it follows that the $n$th-order redundancy $\delta^n_\theta(C^{n,n}) = O(\sqrt{\log n/n})$ for every $\theta \in \Theta$. Then the decomposition
$$
\delta_\theta(C^{n,n}) = \delta^n_\theta(C^{n,n}) + \wh{D}^n_\theta(R) - D_\theta(R)
$$
and (\ref{eq:drf_conv}) imply that (\ref{eq:distred}) holds for every $\theta \in \Theta$.

To prove (\ref{eq:srcident}), fix an $\epsilon > 0$ and note that by (\ref{eq:mindist}), (\ref{eq:2nd_term_noexp}) and the triangle inequality, $d_V(P_\theta,P_{\wh{\theta}}) >
\epsilon$ implies that $2\Delta_\theta(Z^n) + 3/2n + \beta'\sqrt{k/n}  > \epsilon$. Hence,
$$
\Pr \left\{ d_V(P_\theta,P_{\wh{\theta}}) > \epsilon \right\} \le \Pr \left\{ \Delta_\theta(Z^n) > \big(\epsilon - \gamma\sqrt{1/n}\big)/2 \right\},
$$
where $\gamma = 3/2 + \beta'\sqrt{k}$. Since $\cA_\Theta$ is a VC class,
\begin{equation}
\Pr \left\{ d_V(P_\theta,P_{\wh{\theta}}) > \epsilon \right\} \le
8n^{V(\cA_\Theta)} e^{-n(\epsilon - \gamma\sqrt{1/n})^2/128}.
\label{eq:lgdev}
\end{equation}
If for each $n$ we choose $\epsilon_n > \sqrt{128 V(\cA_\Theta) \ln n /n} + \gamma \sqrt{1/n}$, then the
right-hand side of (\ref{eq:lgdev}) will be summable in $n$, hence
$d_V(P_\theta,P_{\wh{\theta}(Z^n)}) = O(\sqrt{\log n/n})$ $P_\theta$-a.s. by the
Borel--Cantelli lemma.\end{proof}
The above proof combines techniques of Rissanen \cite{Ris84} (namely, explicit identification of the source parameters) with the parameter-space quantization idea of Chou {\em et al.} \cite{ChoEffGra96}. The VC condition on the Yatracos class $\cA_\Theta$ is needed to control the $L_1$ convergence rate of the density estimators, which bounds the convergence rate of the distortion redundancies. We remark also that the boundedness condition on the distortion function can be relaxed in favor of a uniform moment condition with respect to a reference letter, but at the expense of a quadratic slowdown of the rate at which the distortion redundancy converges to zero (the proof is omitted for lack of space):

\begin{theorem}Let $\{P_\theta
: \theta \in \Theta\}$ be a family of i.i.d. sources satisfying the conditions of Theorem~\ref{thm:wu}, and let $\rho$ be a distortion function for which
there exists a {\em reference letter} $a_* \in \hcX$ such that $\Sup_{\theta \in \Theta} \E_\theta[\rho(X,a_*)^2] < \infty$, and which satisfies Condition (D.1). Then for any rate $R > 0$ satisfying $\Sup_{\theta \in \Theta} D_\theta(R) < \infty$ there exists a sequence $\{C^{n,n}\}^\infty_{n=1}$ of $(n,n)$-block codes with $R(C^{n,n}) = R + O(\log n/n)$ and $\delta_\theta(C^{n,n}) = O(\sqrt[4]{\log n/n})$ for every $\theta \in \Theta$. The source identification performance is the same as in Theorem~\ref{thm:wu}.
\end{theorem}

\section{Examples}
\label{sec:examples}

Here, we present two explicit examples of parametric families satisfying the conditions of Theorem~\ref{thm:wu} and thus admitting joint universal lossy coding and identification schemes.

\noindent{\bf Mixture classes.} Let $p_1,\cdots,p_k$ be fixed pdf's over a measurable domain $\cX \subseteq \R^d$, and let $\Theta$ be the simplex of probability distributions on $\{1,\cdots,k\}$. Then the mixture class defined
by the $p_i$'s consists of all densities of the form $p_\theta(x) = \sum^k_{i=1} \theta_i p_i(x)$, $\theta = (\theta_1,\cdots,\theta_k) \in \Theta$. The parameter space $\Theta$ is compact and thus satisfies Condition 1) of Theorem~\ref{thm:wu}. In order to show that Condition
2) holds, fix any $\theta,\eta \in \Theta$. Then
$$
d_V(P_\theta,P_\eta) \le  \frac{1}{2}\sum^k_{i=1}|\theta_i - \eta_i| \le \frac{\sqrt{k}}{2} \|\theta - \eta\|,
$$
where the last inequality follows by concavity of the square root. Therefore, the map $\theta \mapsto P_\theta$ is everywhere Lipschitz with
Lipschitz constant $\sqrt{k}/2$. It remains to show that the Yatracos class $\cA_\Theta$ is VC. To this end, observe that
$x \in A_{\theta,\eta}$ if and only if $\sum_i (\theta_i - \eta_i)p_i(x) > 0$. Thus $\cA_\Theta$ consists of sets of the form $\left\{ x \in \cX: \sum_i \alpha_i p_i(x) > 0, (\alpha_1,\cdots,\alpha_k) \in \R^k \right\}$. Since the functions $p_1,\cdots,p_k$ span a linear space of dimension not larger than $k$, Lemma~4.2 in \cite{DevLug01} guarantees that $V(\cA_\Theta) \le k$.

\noindent{\bf Exponential families.} Let $\cX$ be a measurable subset of $\R^d$, and let $\Theta$ be a
compact subset of $\R^k$. A family $\{ p_\theta : \theta \in \Theta\}$
of probability densities on $\cX$ is an {\em exponential family} \cite{BarShe91} if there exist a probability density $p$ and $k$ real-valued functions $h_1,\cdots,h_k$ on $\cX$, such that each $p_\theta$ has the form $p_\theta(x) =  p(x) e^{\theta \cdot h(x) -
g(\theta)}$, where $h(x) \deq
(h_1(x),\cdots,h_k(x))$, $\theta \cdot h(x) \deq \sum^k_{i=1}
\theta_i h_i(x)$, and $g(\theta) = \ln \int_\cX e^{\theta \cdot h(x)}p(x)dx$ is the normalization constant. Given the densities $p$ and $p_\theta$, let $P$ and
$P_\theta$ denote the corresponding distributions. By the compactness of $\Theta$, Condition 1) of
Theorem~\ref{thm:wu} is satisfied. Next we demonstrate that Conditions 2) and 3) can also be met under certain regularity assumptions.

Namely, suppose that $\{1,h_1,\cdots,h_k\}$ is a linearly
independent set of functions. This guarantees that the map $\theta \mapsto P_\theta$ is one-to-one. We also assume that each $h_i$ is
square-integrable with respect to $P$: $\int_\cX h_i^2 dP < \infty$, $1 \le i \le k$. Then the $(k+1)$-dimensional real linear space $\cF \subset L^2(\cX,P)$ spanned by $\{1,h_1,\cdots,h_k\}$ can
be equipped with an inner product $\ave{f,g} \deq \int_\cX fg dP$ and the corresponding $L_2$ norm $\| f \|_2 \deq \sqrt{ \ave{f,f} } \equiv \sqrt{\int_\cX f^2 dP}$. Also let $\| f \|_\infty \deq \inf \big\{ M : |f(x)| \le M \mbox{ $P$-a.e.} \big\}$ denote the $L_\infty$ norm of $f$. Since $\cF$ is
finite-dimensional, there exists a constant $A_k > 0$ such that $\| f \|_\infty \le A_k \| f \|_2$. Finally, assume that the logarithms of Radon--Nikodym derivatives $dP/dP_\theta \equiv p/p_\theta$ are uniformly bounded $P$-a.e.: $\Sup_{\theta \in \Theta} \| \ln (p/p_\theta) \|_\infty = L < \infty$.  These conditions are satisfied, for example, by truncated Gaussian densities over a compact domain in $\R^d$ with suitably bounded means and covariance matrices.

Let $D(P_\theta \| P_\eta)$ denote the relative entropy (information divergence) between $P_\theta$ and $P_\eta$. With the above conditions in place, we can prove the following result along the lines of Lemma~4 of Barron and Sheu \cite{BarShe91}:
\begin{equation}
D(P_\theta \| P_\eta) \le \frac{1}{2}e^{\| \ln (p/p_\theta)
  \|_\infty} e^{2 A_k \| \theta - \eta \|} \| \theta - \eta
  \|^2,
\label{eq:divbound}
\end{equation}
where $\| \cdot \|$ is the Euclidean norm on $\R^k$. From Pinsker's inequality $d_V(P_\theta,P_\eta) \le
\sqrt{D(P_\theta \| P_\eta)/2}$ \cite[Lemma~5.2.8]{Gra90a}, (\ref{eq:divbound}) and the uniform boundedness of $\ln p/p_\theta$, we get
\begin{equation}
d_V(P_\theta,P_\eta) \le \beta_0 e^{A_k\|\theta - \eta \|} \| \theta - \eta \|, \qquad
\theta,\eta \in \Theta,
\label{eq:dvbound}
\end{equation}
where $\beta_0 \deq e^{L/2}/2$. If we fix $\theta \in \Theta$, then from (\ref{eq:dvbound}) it
follows that for any $r > 0$, $d_V(P_\theta,P_\eta) \le \beta_0 e^{A_k r} \| \theta - \eta \|$ for all $\eta$ satisfying $\| \eta - \theta \| \le r$. That is,
the family $\{P_\theta : \theta \in \Theta \}$ satisfies the uniform local
Lipschitz condition of Theorem~\ref{thm:wu}, and the magnitude of the Lipschitz constant can be controlled
by tuning $r$.

All we have left to show is that the Yatracos class $\cA_\Theta$ is a VC class. Since $p_\theta(x) > p_\eta(x)$ if and only if
$(\theta - \eta) \cdot h(x) > g(\theta) - g(\eta)$, $\cA_\Theta$ consists of sets of the form
$$
\Big\{x \in \cX : \alpha_0 + \sum_i \alpha_i h_i(x) > 0, (\alpha_0,\alpha_1,\cdots,\alpha_k) \in \R^{k+1}\Big\}.
$$
Since the functions $1,h_1,\cdots,h_k$ span a $(k+1)$-dimensional
linear space, the same argument as that used for mixture classes shows that $V(\cA_\Theta) \le k+1$.

\section{Conclusion and future work}
\label{sec:conclusion}

We have presented a constructive proof of the existence of a scheme for joint universal lossy block coding and identification of real i.i.d. vector sources with parametric marginal distributions satisfying certain regularity conditions. Our main motivation was to show that the connection between universal coding and source identification, exhibited by Rissanen for lossless coding of discrete-alphabet sources \cite{Ris84}, carries over to the domain of lossy codes and continuous alphabets. As far as future work is concerned, it would be of both theoretical and practical interest to extend the approach described here to variable-rate codes (thus lifting the boundedness requirement for the parameter space) and to general (not necessarily memoryless) stationary sources.

\section*{Acknowledgment}

The author would like to thank Pierre Moulin and Ioannis Kontoyiannis for useful discussions. This work was supported by the Beckman Institute Postdoctoral Fellowship. 

\bibliography{ucm_isit2006}

\begin{thebibliography}{10}
\providecommand{\url}[1]{#1}
\csname url@rmstyle\endcsname
\providecommand{\newblock}{\relax}
\providecommand{\bibinfo}[2]{#2}
\providecommand\BIBentrySTDinterwordspacing{\spaceskip=0pt\relax}
\providecommand\BIBentryALTinterwordstretchfactor{4}
\providecommand\BIBentryALTinterwordspacing{\spaceskip=\fontdimen2\font plus
\BIBentryALTinterwordstretchfactor\fontdimen3\font minus
  \fontdimen4\font\relax}
\providecommand\BIBforeignlanguage[2]{{%
\expandafter\ifx\csname l@#1\endcsname\relax
\typeout{** WARNING: IEEEtran.bst: No hyphenation pattern has been}%
\typeout{** loaded for the language `#1'. Using the pattern for}%
\typeout{** the default language instead.}%
\else
\language=\csname l@#1\endcsname
\fi
#2}}

\bibitem{Ris84}
J.~Rissanen, ``Universal coding, information, prediction, and estimation,''
  \emph{IEEE Trans. Inform. Theory}, vol. IT-30, no.~4, pp. 629--636, July
  1984.

\bibitem{NeuGraDav75}
D.~L. Neuhoff, R.~M. Gray, and L.~D. Davisson, ``Fixed rate universal block
  source coding with a fidelity criterion,'' \emph{IEEE Trans. Inform. Theory},
  vol. IT-21, no.~5, pp. 511--523, September 1975.
  
  \bibitem{LinLugZeg94}
T.~Linder, G.~Lugosi, and K.~Zeger, ``Rates of convergence in the source coding
  theorem, in empirical quantizer design, and in universal lossy source
  coding,'' \emph{IEEE Trans. Inform. Theory}, vol.~40, no.~6, pp. 1728--1740,
  November 1994.

\bibitem{ChoEffGra96}
P.~A. Chou, M.~Effros, and R.~M. Gray, ``A vector quantization approach to
  universal noiseless coding and quantization,'' \emph{IEEE Trans. Inform.
  Theory}, vol.~42, no.~4, pp. 1109--1138, July 1996.

\bibitem{MouKoe05}
P.~Moulin and R.~Koetter, ``Data-hiding codes,'' \emph{Proc. IEEE}, vol.~93,
  no.~12, pp. 2085--2127, December 2005.

\bibitem{DevLug01}
L.~Devroye and G.~Lugosi, \emph{Combinatorial Methods in Density
  Estimation}.\hskip 1em plus 0.5em minus 0.4em\relax New York:
  Springer-Verlag, 2001.
  
\bibitem{CsiKor81}
I.~Csisz\'ar and J.~K\"orner, \emph{Information Theory: Coding Theorems for
  Discrete Memoryless Sources}.\hskip 1em plus 0.5em minus 0.4em\relax
  Budapest: Akad\'emiai Kiad\'o, 1981.

\bibitem{Yat85}
Y.~G. Yatracos, ``Rates of convergence of minimum distance estimates and
  {K}olmogorov's entropy,'' \emph{Ann. Math. Statist.}, vol.~13, pp. 768--774,
  1985.

\bibitem{DevLug96}
L.~Devroye and G.~Lugosi, ``A universally acceptable smoothing factor for
  kernel density estimation,'' \emph{Ann. Statist.}, vol.~24, pp. 2499--2512,
  1996.

\bibitem{DevLug97}
------, ``Nonasymptotic universal smoothing factors, kernel complexity and
  {Y}atracos classes,'' \emph{Ann. Statist.}, vol.~25, pp. 2626--2637, 1997.

\bibitem{BarShe91}
A.~R. Barron and C.-H. Sheu, ``Approximation of density functions by sequences
  of exponential families,'' \emph{Ann. Statist.}, vol.~19, no.~3, pp.
  1347--1369, 1991.

\bibitem{Gra90a}
R.~M. Gray, \emph{Entropy and Information Theory}.\hskip 1em plus 0.5em minus
  0.4em\relax New York: Springer-Verlag, 1990.





\end{thebibliography}

\end{document}